\documentclass[aps,prl,secnumarabic,nobibnotes,twocolumn,superscriptaddress,longbibliography]{revtex4-1}

\usepackage{amsfonts}
\usepackage{amsmath}
\usepackage{color}
\usepackage{graphicx}
\usepackage{bm}

\usepackage{xspace}
\usepackage{epstopdf}
\usepackage{dcolumn}
\usepackage{longtable}
\usepackage{multirow}
\usepackage[colorlinks=true, letterpaper=true, pdfstartview=FitV, linkcolor=blue, citecolor=blue, urlcolor=blue]{hyperref}

\makeatother
\begin{document}
\title{Quantized Circulation of Anomalous Shift in Interface Reflection}

\author{Ying Liu}
\affiliation{School of Materials Science and Engineering, Hebei University of Technology, Tianjin 300130, China}

\affiliation{Research Laboratory for Quantum Materials, Singapore University of
Technology and Design, Singapore 487372, Singapore}

\author{Zhi-Ming Yu}
\email{zhiming\_yu@bit.edu.cn}
\affiliation{Key Lab of Advanced Optoelectronic Quantum Architecture and Measurement (MOE),
Beijing Key Lab of Nanophotonics \& Ultrafine Optoelectronic Systems,
and School of Physics, Beijing Institute of Technology, Beijing 100081, China}
\affiliation{Research Laboratory for Quantum Materials, Singapore University of
Technology and Design, Singapore 487372, Singapore}

\author{Cong Xiao}
\affiliation{Department of Physics, The University of Texas at Austin, Austin, Texas 78712, USA}

\author{Shengyuan A. Yang}
\affiliation{Research Laboratory for Quantum Materials, Singapore University of
Technology and Design, Singapore 487372, Singapore}
\affiliation{Center for Quantum Transport and Thermal Energy Science,
School of Physics and Technology, Nanjing Normal University, Nanjing 210023, China}

\begin{abstract}
A particle beam may undergo an anomalous spatial shift when it is reflected at an interface. The shift forms a vector field defined in the two-dimensional interface momentum space. We show that, although the shift vector at individual momentum is typically sensitive to the system details, its integral along a close loop, i.e., its circulation, could yield a robust quantized number under certain conditions of interest. Particularly, this is the case when the beam is incident from a trivial medium, then the quantized circulation of anomalous shift (CAS)
directly manifests the topological character of the other medium. We demonstrate that the topological charge of a Weyl medium as well as
the unconventional pair potentials of a superconductor can be captured and distinguished by CAS. Our work unveils a hidden quantized feature in a ubiquitous physical process, which may also offer a new approach for probing topological media.
\end{abstract}

\maketitle

Quantized quantities are rare and always fascinating in physics. Such quantities, e.g., the Hall conductivity in quantum Hall effect~\cite{Klitzing1980} and the circulation in superfluid~\cite{Onsager1949a}, are invariably considered as remarkable, because they reveal deep physics and allow the rare chance of high precision measurement. In this work, we reveal a new member in an ubiquitous physical process: the interface reflection.

\begin{figure}[t!]
\includegraphics[width=8.5cm]{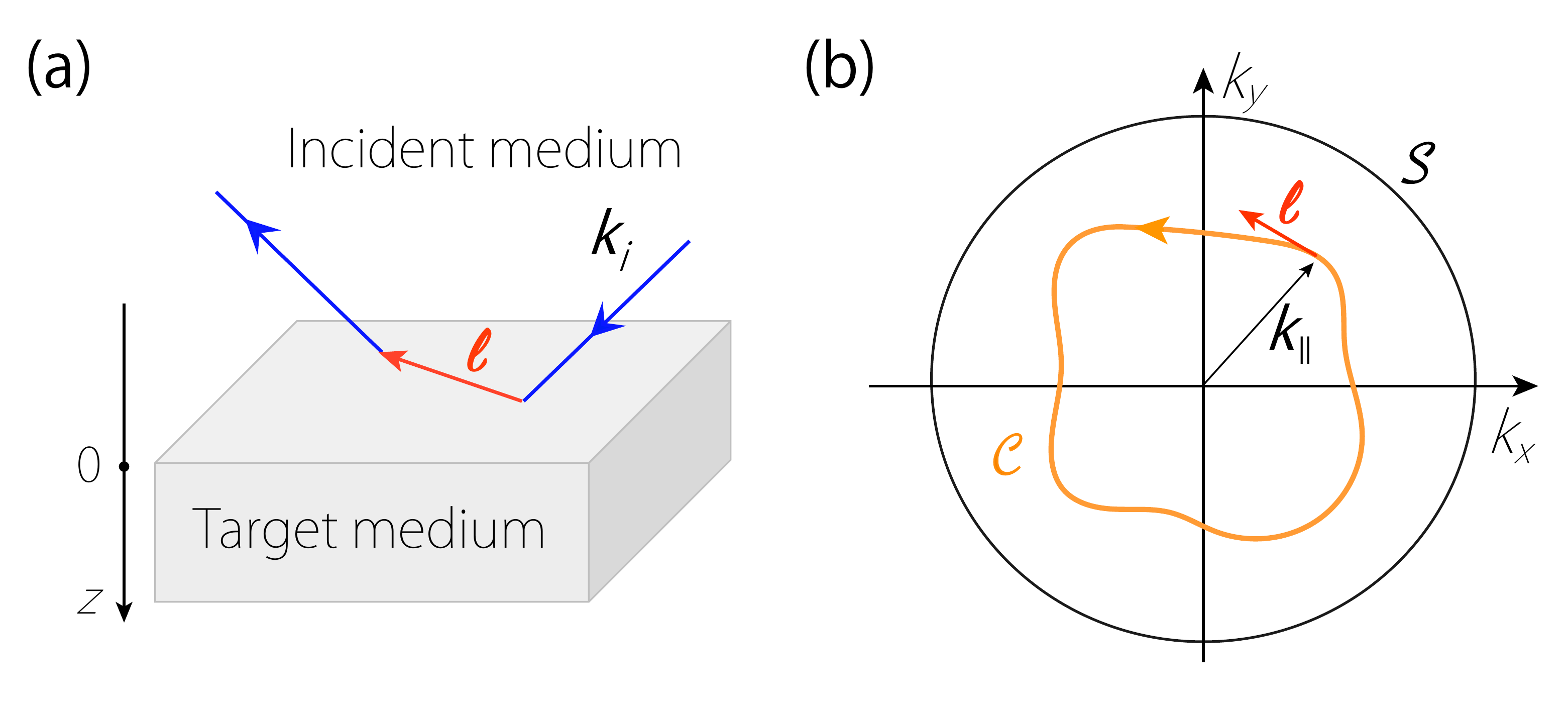}\caption{(a) Schematic figure showing the anomalous spatial shift $\bm\ell$ for a beam reflected at an interface. (b) $\bm \ell$ forms a vector field in the domain $S$ (where reflection occurs) in the interface momentum $\bm k_\|$-space. The CAS $\kappa_C$ is defined for any closed loop $C\in S$. \label{fig: 1}}
\end{figure}

In the simplest picture, a light beam reflected at a flat sharp interface should follow the law of reflection, which assumes that the reflected and the incident beams meet at the same point on the interface. However, the wave nature of light brings a twist to this simple picture: the reflected beam may acquire an anomalous spatial shift from the incident point, as illustrated in Fig.~\ref{fig: 1}(a)~\cite{angular}. The longitudinal and the transverse components of this shift (defined with respect to the incident plane) represent the well-known Goos-H\"{a}nchen effect~\cite{Goos1947} and Imbert-Fedorov effect~\cite{Fedorov1955,Imbert1972}, respectively. Remarkably, the analogous effects have also been discovered in electronic scattering  \cite{Miller1972,Fradkin1974,Sinitsyn2005,Chen2008,BeenakkerGH2009,Sharma2011,WuPRL2011,Chen2011,JiangPRL2015,YangPRL2015,JiangPRB2016,WangPRB2017,Chattopadhyay2018}, and most recently in Andreev reflection \cite{LiuSUTDPRB2017,LiuCAR2018,YuSUTDPRL2018}. For the latter, the shift occurs at a normal-metal/superconductor interface, between the incident electron beam and the reflected hole beam.
These examples demonstrate that the effect is general for both classical and quantum systems, so it has been attracting broad interest~\cite{OnodaPRL2004,BliokhPRL2006,BliokhNPhoto2015,KKWJMprb2016,YUFOP2019}.

Here, we unveil a hidden quantized feature of this general effect. We construct a character, termed as the circulation of anomalous shift (CAS), which is the integral of the shift vector along a closed path in the interface momentum space. We show that for \emph{incident} medium satisfying certain symmetries, CAS must take a quantized value, analogous to the circulation in superfluid. This quantization applies for a class of interesting cases. For example, we consider the shift for a beam incident from a trivial medium onto the interface with a Weyl medium, and show that the quantized CAS characterizes the topological charge of the Weyl point. As another example, we show that the CAS for Andreev reflection at the interface between a simple metal and a superconductor can distinguish different superconducting pairing types. The quantized feature makes CAS a robust topological quantity against perturbations. The findings here also provide a new approach for probing nontrivial topological or superconducting states.

\textit{\textcolor{blue}{Anomalous shift in interface scattering}}. We start by presenting a general formula for this shift.
Consider the setup in Fig.~\ref{fig: 1}(a), where a flat interface between two media is formed at the $z=0$ plane. A beam of particles $\bm\Psi^i_{\bm k^i}$ incoming from the upper medium (referred to as the \emph{incident medium}) is scattered at the interface into a reflected beam $\bm\Psi^r_{\bm k^r}$ and a transmitted beam $\bm\Psi^t_{\bm k^t}$.
The $\bm\Psi$'s are usually modeled by wave packets, and are required to be confined in both real and momentum spaces. $\bm k^\alpha$ ($\alpha=i, r, t$) denote their average momenta, and must share the same conserved component $\bm k_\|$ in the interface ($x$-$y$) plane.
To analyze the scattering, the beam is expanded using the scattering eigenstates. For instance, $\bm\Psi^i_{\bm k^i}=\int d\bm k'w(\bm k'-\bm k^i)\psi^i_{\bm k'}$, where $w$ is the profile of the beam peaked at $\bm k^i$, and $\psi^i_{\bm k}=e^{i\bm k\cdot\bm r}|u^i_{\bm k}\rangle$ is the Bloch eigenstate of the incident medium with $|u^i_{\bm k}\rangle$ being the cell-periodic part. The scattering of each partial wave $\psi^i$ into reflected wave $\psi^r$ and transmitted wave $\psi^t$ is captured by the scattering amplitudes $r$ and $t$, respectively. Hence, the reflected beam can be expressed as $\bm\Psi^r_{\bm k^r}=\int d\bm k'w(\bm k'-\bm k^r)r(\bm k')\psi^r_{\bm k'}$, similar for the transmitted one.
The anomalous shift is found by comparing
the center positions of the beams at the interface.

Following the standard approach~\cite{BeenakkerGH2009,JiangPRL2015,LiuSUTDPRB2017,YUFOP2019}, the spatial shift for the reflected/transmitted beam can be obtained as \cite{ShiPRB2019}
\begin{equation}\label{shift}
  \bm \ell^s=\left\langle u^s_{\bm k^s}\left|i\frac{\partial}{\partial \bm k_\|}\right|u^s_{\bm k^s}\right\rangle
  -\left\langle u^i_{\bm k^i}\left|i\frac{\partial}{\partial \bm k_\|}\right|u^i_{\bm k^i}\right\rangle-\frac{\partial}{\partial \bm k_\|}\text{arg}(s),
\end{equation}
where $s=r,t$, and in the last term, we have abused the superscript $s$ to also denote the scattering amplitude.

In this formula, the first two terms each is the in-plane component of the Berry connection $\bm{\mathcal{A}}=\langle u|i\nabla_{\bm k}|u\rangle$ for the corresponding state, which is an intrinsic band geometric property. The last term shows that the shift depends on the phase but not the magnitude of the scattering amplitude. In Fig.~\ref{fig: 1}(a), we considered a single reflected/transimitted beam. The result also applies when there are multiple scattering channels, simply by inserting the state and the scattering amplitude for the corresponding beam $s$.
We note that the formula closely resembles the result for the side jump at an impurity derived by Sinitsyn, Niu, and MacDonald \cite{SinitsynPRB2006}.

\textit{\textcolor{blue}{CAS and its quantization}}. Now let us proceed to the concept of CAS. For concreteness, we focus on the shift for the reflected beam and neglect the superscript $r$ in the following discussion.

As from Eq.~(\ref{shift}), the shift $\bm \ell$ is a function of the interface momentum $\bm k_\|$, which is conserved during scattering. Assuming the equi-energy surface in the incident medium takes a simple convex shape, $\bm \ell$ forms a vector field defined in a domain $S$ (where reflection occurs) in the interface momentum $\bm k_\|$-space [Fig.~\ref{fig: 1}(b)]. Then the CAS $\kappa_C$ along a closed loop $C\in S$ is defined as
\begin{equation}
  \kappa_C\equiv \oint_C\bm \ell\cdot d\bm k_\|.
\end{equation}
Clearly, for a generic loop, both longitudinal and  transverse components of $\bm{\ell}$ contribute to the CAS.
Using formula~(\ref{shift}), $\kappa_C$ can be expressed as two contributions
\begin{equation}
  \kappa_C=\Delta\gamma-\oint_C\frac{\partial}{\partial \bm k_\|}\text{arg}(r)\cdot d\bm k_\|.
\end{equation}
Here, the first contribution $\Delta \gamma$ stands for the integral of the first two terms along $C$. Importantly, the second contribution represents the phase winding of the reflection amplitude. As long as the loop does not hit any singularity of $\text{arg}(r)$, this term must give an integer multiple of $2\pi$, namely, we should have
\begin{equation}
  \kappa_C=\Delta\gamma+2\pi N,\qquad N\in\mathbb{Z}.
\end{equation}

Due to the $\Delta\gamma$ term, $\kappa_C$ need not be quantized for the most general case. However, there are many cases of interest, where $\Delta\gamma$ vanishes and we indeed have a quantized CAS. This is true (at least) when either of the following conditions is satisfied. (i) The incident medium is ``trivial", in the sense that its Berry connection vanishes. This is the case when the medium has a real representation (i.e., described by a real Hamiltonian). If satisfied, $\Delta\gamma$ must vanish, and $\kappa_C=2\pi N$.
(ii) The incident medium has a reflection symmetry $\mathcal{M}_z$ which connects the incident and the reflected states (this mirror must be parallel to the interface). In this case, even though the Berry connections for the incident and the reflected states may be nonzero, their in-plane components must cancel out and hence $\Delta\gamma=0$.

Under the above conditions, the CAS takes a quantized value, analogous to the circulation in superfluid~\cite{Onsager1949a}. Importantly, the quantized CAS is solely determined by the scattering amplitude $r$, or more specifically, by the phase winding of $r$, which encodes the information of the other medium at $z>0$ (referred to as the \emph{target medium}). A nontrivial CAS value ($N\neq 0$) indicates generically
the presence of vortices in the vector field $\bm \ell$, and the locations of these vortices correspond to the singularities in the phase of $r$.
As we shall see, the CAS can manifest the distinct features of the target medium. Moreover, the quantization of CAS endows it with a topological robustness, namely, its value is robust against perturbations on the system. Thus, the quantized CAS provides a powerful way to characterize medium properties.

To illustrate these points, we consider two concrete examples below.

\begin{figure}[ht]
\includegraphics[width=8.5cm]{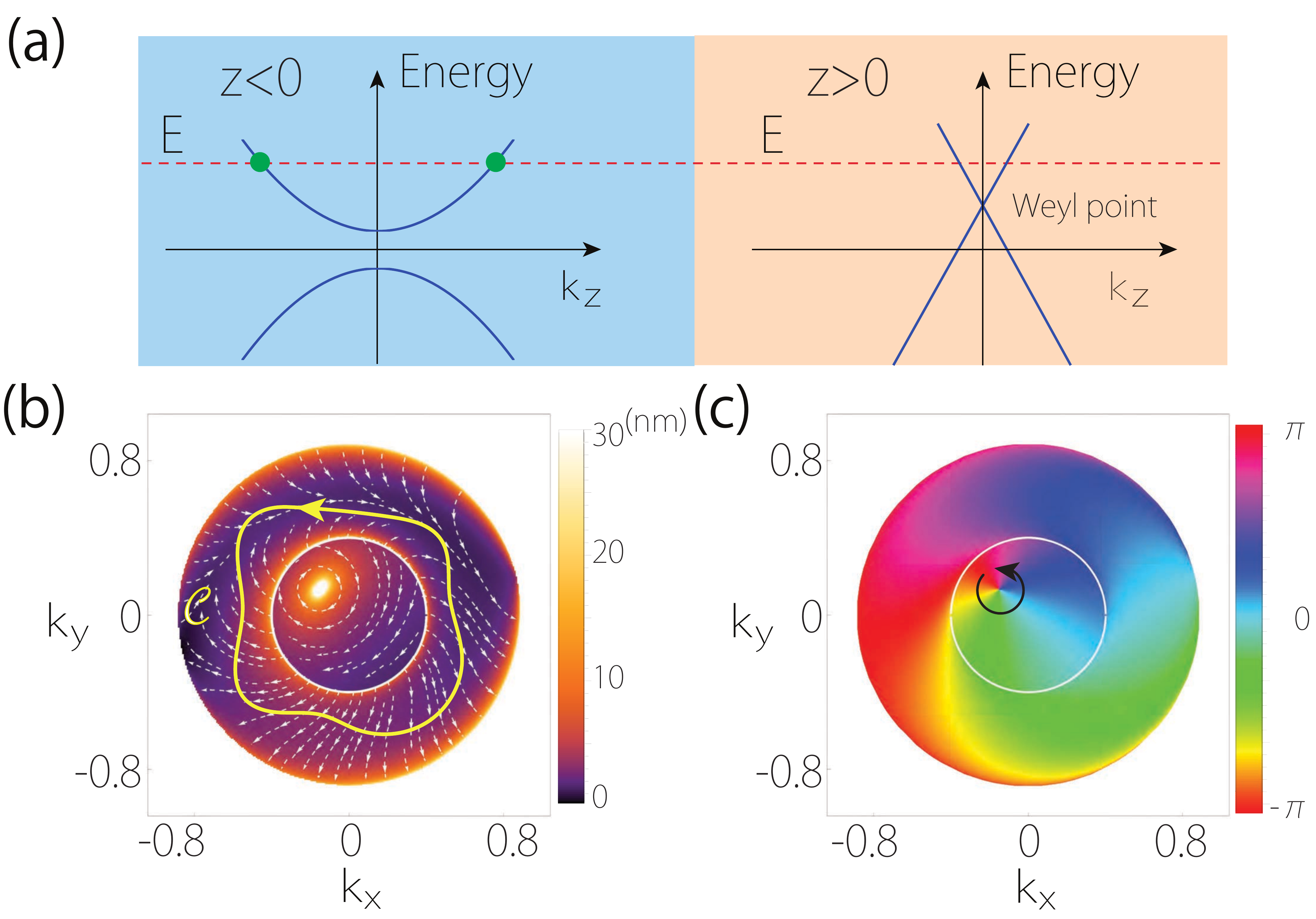}\caption{(a) Schematic figure showing the interface formed between a trivial medium and a Weyl medium. (b)
Numerical results for the $\bm \ell$ field, for which the magnitude and the direction are indicated by the color and the streamline,
respectively. (c) Plot of $\text{arg}(r)$. The white circles in (b) and (c) indicate the region with radius $k_W$.
Here, we set $m=0.04m_e$, $\Delta=0.2$ eV, $v=1.5\times 10^{6}$ m/s, $U=0.6$ eV, $E=1.0$ eV, and $\delta=10^{-4}$ eV$\cdot$nm$^2$. \label{fig: 2}}
\end{figure}

\textit{\textcolor{blue}{Probing a Weyl medium}}. In the first example, we take a trivial incident medium ($z<0$) described by a simple two-band semiconductor-like model
\begin{eqnarray}
H_\text{I} & = & \left(\frac{k^{2}}{2m}+\Delta\right)\sigma_{z},
\end{eqnarray}
where $2\Delta$ is the gap between the two bands, and the Pauli matrix $\sigma$ here stands for a pseudospin (e.g., orbital) degree of freedom.
The target medium ($z>0$) that we try to probe is a Weyl medium \cite{Wan2011,ArmitageRMP2018}, described by
\begin{eqnarray}\label{WeylH}
H_\text{T} & = & vk_{z}\sigma_{z}+w\left(k_{-}^{n}\sigma_{+}+k_{+}^{n}\sigma_{-}\right)+U+\delta k^2 \sigma_x,
\end{eqnarray}
where $k_{\pm}=k_{x}\pm ik_{y}$, $\sigma_{\pm}=(\sigma_{x}\pm i\sigma_{y})/2$, $n$ is a positive integer, $v$, $w$, $U$, and $\delta$ are model parameters. The first two terms describe a Weyl point at the origin with a topological charge
\begin{equation}
  \nu=\text{sgn}(v)n.
\end{equation}
For the case with $n=1$ and $w=v$, they reduce to the usual Weyl model $v\bm k\cdot \bm \sigma$; and the case with $n>1$ corresponds to the so-called multi-Weyl point with higher topological charges \cite{FangPRL2012,Xu2011}. $U$ represents a potential energy shift across the interface. The last term is a small quadratic term added to ensure a well-defined boundary condition with $H_\text{I}$, which does not affect the essential physics, and we may take $\delta\rightarrow 0$ in the final results.

Because $H_\text{I}$ is trivial, according to our theory, the CAS must be quantized. More importantly, as we shall see, the CAS can take a nonzero value, which is determined by the topological charge $\nu$ of the Weyl point inside the target medium.

Let us first consider the conventional Weyl point with $n=1$. Without loss of generality, we assume the incident beam has energy $E>\Delta$, i.e., in the upper band. Our discussion of CAS is within the domain $S$ with $k_\|<k_M$, where $k_M=\sqrt{2m (E-\Delta)}$ is the maximal interface momentum at $E$. For each $\bm k_\|\in S$, the reflection amplitude $r$ and the shift $\bm \ell$ can be directly calculated using the standard approach~\cite{SI}.

Here, to probe the Weyl point, we are most interested in the situation when the Weyl point energy is close to $E$, as illustrated in Fig.~\ref{fig: 2}(a). In such a case, the equi-energy surface of $E$ in the Weyl medium has a radius $k_W=|(E-U)/w|< k_M$ in the $k_x$-$k_y$ plane, so its projection in the $\bm k_\|$-plane lies within $S$. Now, consider the annulus-shaped region $A\subset S$ with $k_W<k_\|<k_M$. Within $A$, we must have total reflection with $|r|^2=1$, because there is no transmitted state in the Weyl medium with such $\bm k_\|$. This indicates that $\text{arg}(r)$ has no singularity in $A$, and $\bm \ell$ is analytic in this region. Therefore, for any two topologically equivalent loops $C_1, C_2\in A$, we must have $\kappa_{C_1}=\kappa_{C_2}$.

As any contractible loop in $A$ must have $\kappa=0$, let us consider a non-contractible loop $C\in A$ that encircles the inner hole of the annulus [see Fig.~\ref{fig: 2}(b)]. According to our analysis, the calculation of $\kappa_C$ can be simplified by noting that $\kappa_C=\kappa_{C_0}$, where $C_0$ is a regular circle with radius $k_\|$ in $A$. Through straightforward calculations~\cite{SI}, we obtain the transverse component of the shift $\ell_\phi(\equiv \bm\ell\cdot\hat{\phi})$ ,
\begin{equation}
  \ell_\phi=\frac{1-|\zeta|^2}{k_\|\times|\zeta+i\eta e^{-i\phi}|^2},
\end{equation}
where $\phi=\text{arg}(k_x+ik_y)$ is the polar angle of $\bm k_\|$, {{$\zeta=\frac{(p_-+mv)(mv+ip_+)}{(p_--mv)(mv-ip_+)}$, $\eta=\mathrm{sgn}(E-U)(\frac{E-U-vp_0}{E-U+vp_0})^{1/2}$}}, with $p_\pm=[2m(E\pm\Delta)\pm k_\|^2]^{1/2}$ and {{$p_0=\mathrm{sgn}(E-U)[(E-U)^2-w^2k_\|^2]^{1/2}/v$}}. With this result, we obtain
\begin{equation}\label{loop}
  \kappa_C=\kappa_{C_0}=k_\|\int_0^{2\pi}\ell_\phi d\phi=-2\pi\nu.
\end{equation}
This remarkable result shows that an \emph{arbitrary} loop enclosing the hole (i.e., the equi-energy surface of the Weyl medium) has a nontrivial quantized CAS, which is solely determined by the topological charge $\nu$ of the Weyl point. Moreover, this result is independent of whether the Weyl point is above or below the energy $E$.

\begin{figure}[t]
\includegraphics[width=8.5cm]{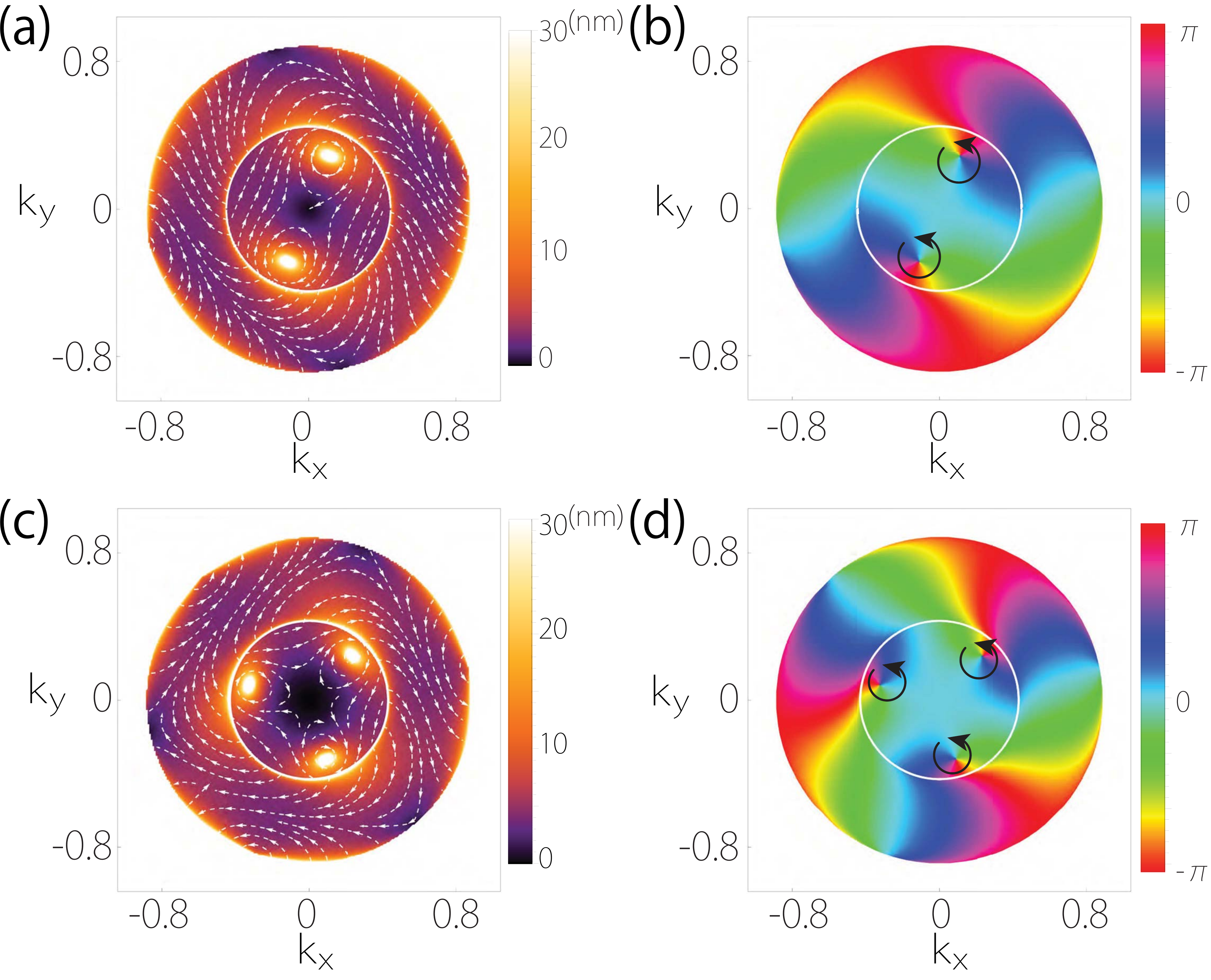}\caption{Results for the target medium with (a-b)
double-Weyl point, and (c-d) triple-Weyl point, corresponding to $n=2$ and $n=3$ in model (\ref{WeylH}), respectively. (a, c) exhibit the $\bm \ell$ field, and (b, d) show the phase of $r$. The white circles indicate the region with radius $k_W$.  Here, we take $w=1$ $\mathrm{nm}^2\cdot\mathrm{eV}$, $U=0.8$ eV in (a-b); and $w=1$ $\mathrm{nm}^3\cdot\mathrm{eV}$, $U=0.92$ eV in (c-d). Other parameters are the same as that in Fig.~\ref{fig: 2}. \label{fig: 3}}
\end{figure}

The nontrivial CAS in Eq.~(\ref{loop}) indicates the existence of vortex in the phase winding of $r$ inside the central region with $k_\|<k_W$. The vortex center corresponds to the singularity in $\text{arg}(r)$, which occurs here at $r=0$. In Fig.~\ref{fig: 2}(c), we plot the numerical result for $\text{arg}(r)$, which confirms the presence of a vortex, and the vortex center coincides with the zero-point of $r$. In the limit of $E\gg \Delta, (E-U)$, we can derive an analytic expression for the shift around the vortex center $\bm k_\|^*$, which exhibits the behavior~\cite{SI}
\begin{equation}
  \bm \ell \propto \nu (q_y, -q_x)/q^2,
\end{equation}
where  $\bm q$ is the momentum measured from $\bm k_\|^*$. This confirms the nontrivial circulation pattern around the vortex. In addition, one can easily see that the result in (\ref{loop}) applies to any loop $C\in S$ that encloses the point $\bm k_\|^*$.

The above analysis can be directly extended to multi-Weyl points with $n>1$.
As spin- or pseudospin-orbit coupling plays an important role in the shift~\cite{YangPRL2015,OnodaPRL2004,LiuSUTDPRB2017}, the $\bm{\ell}$-field patterns for the multi-Weyl points are expected to be different, due to their different forms of coupling.
However, the key finding is that our result (\ref{loop}) remains valid \cite{SI}, namely, the CAS is nontrivial, and the quantization integer $-\kappa_C/(2\pi)$ just corresponds to the topological charge of the Weyl point. In Fig.~\ref{fig: 3}, we plot the calculation results for double-Weyl ($n=2$) and triple-Weyl ($n=3$) points. One observes that there exist $n$ vortices inside the $k_\|<k_W$ region, which are responsible for the nontrivial quantized CAS.

\textit{\textcolor{blue}{Probing a superconductor}}. As a second example, we consider the CAS for Andreev reflection at a normal-metal/superconductor interface. Here, we take the model of a simple metal for the incident medium $(z<0)$, so its Bogoliubov-de Gennes (BdG) Hamiltonian  is given by~\cite{Gennes1966}
\begin{equation}
  H_\text{I}=\left(\frac{k^{2}}{2m}-E_{F}\right)\tau_{z},
\end{equation}
where $E_F$ is the Fermi energy and $\tau$ is the Nambu pseudospin acting on the electron-hole space. The target medium ($z>0$) is a superconductor described by \cite{Gennes1966,Blonder1982,Kashiwaya2000}
\begin{equation}
  H_\text{T}=\left(\frac{k^{2}}{2m}+U-E_{F}\right)\tau_{z}+(\boldsymbol{\Delta}\tau_{+}+\boldsymbol{\Delta}^{*}\tau_{-}),
\end{equation}
where $\boldsymbol{\Delta}$ is the superconducting pair potential and $\tau_\pm=(\tau_x\pm i\tau_y)/2$.

Here, we focus on the shift $\bm \ell$ during Andreev reflection for an incident electron beam with excitation energy $\varepsilon$. Again, because the incident medium ($H_\text{I}$) is trivial, the shift is solely determined by the Andreev-reflection amplitude $r_A$, and its CAS must be quantized. In the following, we shall investigate how the different pairing types (encoded in $\boldsymbol{\Delta}$) affect the value of CAS.

\begin{figure}[t!]
\includegraphics[width=8.5cm]{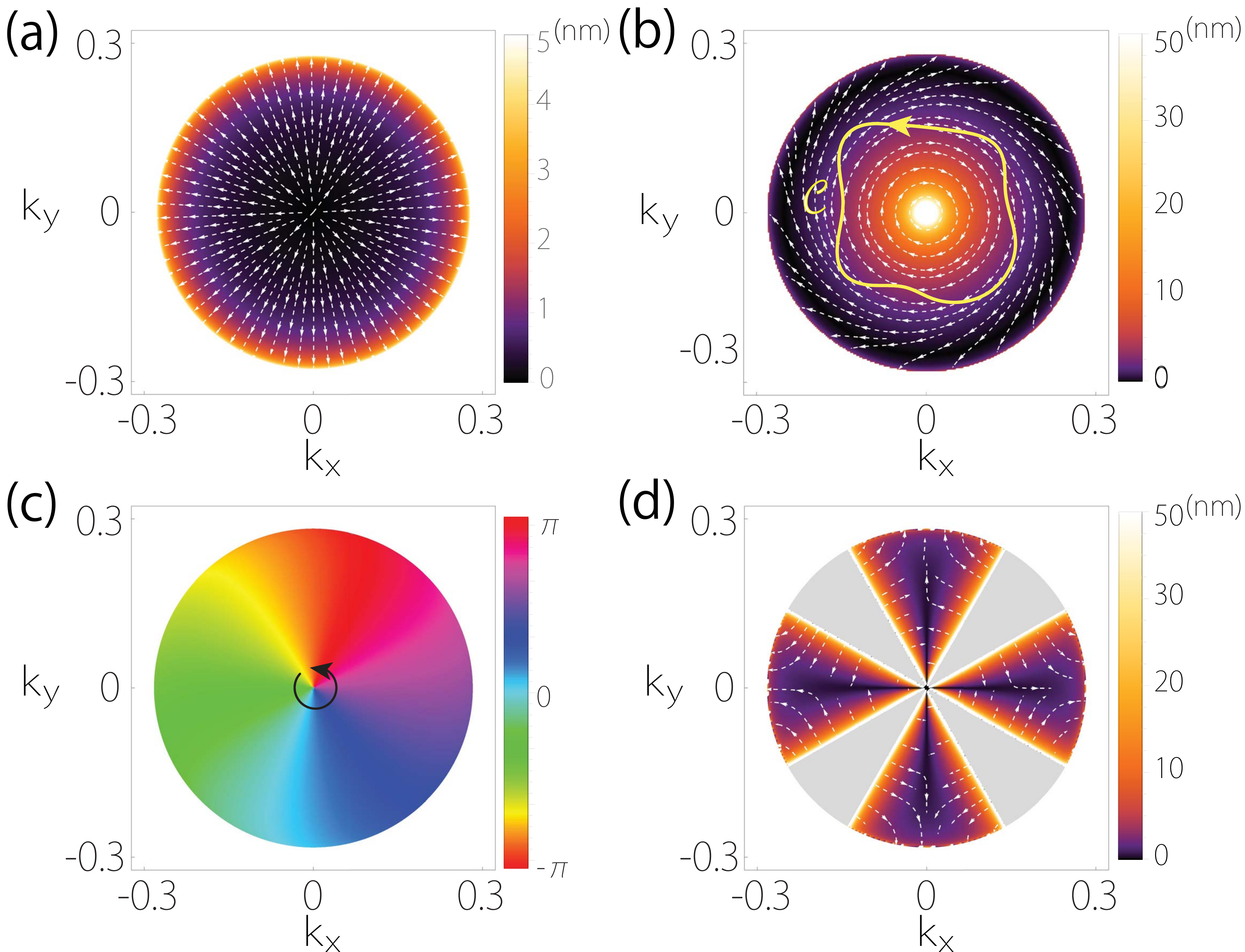}\caption{Calculated $\bm \ell$ field in Andreev reflection for (a) $s$-wave, (b) chiral
$p$-wave, and (d) $d_{x^{2}-y^{2}}$-wave pair potentials. (c) shows the phase of $r_A$ for the chiral $p$-wave case. Here, we take $\varepsilon=0.02$ eV,
$m=0.04 m_e$, $\Delta_{0}=0.04$ eV, $E_{F}=0.1$ eV, and $U=0.3$ eV. \label{fig: 4}}
\end{figure}

We first consider the $s$-wave pair potential with $\boldsymbol{\Delta}$ given by a constant $\Delta_0$. Straightforward calculation~\cite{SI} shows that
the shift only has a longitudinal component [Fig.~\ref{fig: 4}(a)], with
\begin{equation}\label{swave}
  \bm\ell=\frac{2(p_s^2-p_n^2)^2\tan\alpha}{p_s p_n [4  p_s^2 p_n^2+(p_s^2+p_n^2)^2\tan^2\alpha]}\bm k_\|,
\end{equation}
where $p_n=(2m E_F-k_\|^2)^{1/2}$, $p_s=[2m(E_F-U)-k_\|^2]^{1/2}$, and $\alpha=-i\cosh^{-1}(\varepsilon/\Delta_0)$. Meanwhile, one can check that $\text{arg}(r_A)$ has no singularity within $S$. Thus, the CAS vanishes in this case.

The situation is dramatically different for chiral pair potentials, described by $\boldsymbol{\Delta}=\Delta_{0}e^{-i\chi\phi}$ with $\chi$ a nonzero integer. For this case, $\bm\ell$ has not only a longitudinal component [given by the same expression as (\ref{swave})], but also a transverse component
\begin{equation}
  \ell_\phi= -\frac{\chi}{k_\|}.
\end{equation}
One immediately notes that the CAS is nontrivial for any simple loop $C$ encircling the origin  [see Fig. \ref{fig: 4}(b)] and
\begin{equation}\label{chiral}
  \kappa_C=-2\pi \chi,
\end{equation}
i.e., the quantized CAS is solely determined by the chirality of the pairing. Figure~\ref{fig: 4}(c) further confirms the vortex in the phase winding of $r_A$. Note that different from the case in Fig.~\ref{fig: 2}, the singularity here is due to the fact that $\phi$ in the pair potential is undefined at $\bm k_\|=0$.

Finally, we consider the $d_{x^{2}-y^{2}}$-type pairing, with $\boldsymbol{\Delta}=\Delta_{0}\cos(2\phi)$. Figure~\ref{fig: 4}(d) shows the numerical result for $\bm\ell$, which generally has both longitudinal and transverse components. In the figure, the four gray-colored sectors mark the so-called suppressed zones~\cite{YuSUTDPRL2018}, corresponding to the four nodes of the $d$-wave pairing gap, where $\varepsilon>|\boldsymbol{\Delta}|$ and the shift vanishes~\cite{SI}. Excluding the suppressed zones from $S$, then within $S$ there is no isolated singular point. Thus, for a loop $C\in S$, we should have $\kappa_C=0$.

\textit{\textcolor{blue}{Discussion}}.
We reveal a new quantized quantity in an ubiquitous process. Moreover, the quantized CAS encodes the topological information of the medium, therefore may offer a new approach for characterization. As we have shown, this could be particularly useful by constructing a planar interface between a trivial incident medium (such as the simple metal) and the target medium (to be probed), and by mapping out the shift vector field.
This actually conforms with the typical experimental setup. Note that this setup is distinct from the previous works on shift in Weyl semimetals~\cite{JiangPRL2015,YangPRL2015,WangPRB2017,Chattopadhyay2018}, where the beam is incident from a Weyl semimetal, so the incident medium itself is nontrivial and does not satisfy the condition for the CAS quantization. For example, Ref.~\cite{Chattopadhyay2018} reported a kind of ``half" vortex in the $\bm{\ell}$ field, which is not quantized.

As quantized quantities, the CAS and the associated vortices enjoy a topological robustness (the vortices are well-known topological defects of the vector field). For example, one can show that they are robust against possible potential barrier at the interface~\cite{SI}.

The experimental detection of the anomalous shift has been well developed in the field of optics~\cite{Hosten2008,BliokhNPhoto2008,Haefner2009,Herrera2010,Yin2013,ZhouPRA2013}. Photonic crystals simulating Weyl and other topological band structures have also been achieved~\cite{Lu6222015,Yang1013Sci,XueNM2019,YangNP2019}. Hence, the circulation pattern, the vortex, and the quantized CAS should be easily probed in the optical context. Meanwhile, several methods for detecting the shift in electronic systems were also proposed, such as by engineering specific junction geometry~\cite{YangPRL2015,LiuSUTDPRB2017,LiuSUTD2018} and by enhancing the shift via multiple reflections~\cite{JiangPRL2015,YuSUTDPRL2018}. The shift there can typically reach $10\sim 100$ nm~\cite{LiuSUTD2018,YuSUTDPRL2018}, much larger than the lattice scale. It is particularly interesting to probe the shift for superconductors (in Fig.~\ref{fig: 4}), which can help to characterize unconventional superconductivity.

\begin{acknowledgments}
\textit{Acknowledgments---}The authors thank D. L. Deng for helpful discussions. This work is supported by the Singapore Ministry of Education AcRF Tier 2 (MOE2017-T2-2-108) and Beijing Institute of Technology Research Fund Program for Young Scholars.
\end{acknowledgments}


\bibliographystyle{apsrev4-1}
\bibliography{QV}

\begin{thebibliography}{50}%
\makeatletter
\providecommand \@ifxundefined [1]{%
 \@ifx{#1\undefined}
}%
\providecommand \@ifnum [1]{%
 \ifnum #1\expandafter \@firstoftwo
 \else \expandafter \@secondoftwo
 \fi
}%
\providecommand \@ifx [1]{%
 \ifx #1\expandafter \@firstoftwo
 \else \expandafter \@secondoftwo
 \fi
}%
\providecommand \natexlab [1]{#1}%
\providecommand \enquote  [1]{``#1''}%
\providecommand \bibnamefont  [1]{#1}%
\providecommand \bibfnamefont [1]{#1}%
\providecommand \citenamefont [1]{#1}%
\providecommand \href@noop [0]{\@secondoftwo}%
\providecommand \href [0]{\begingroup \@sanitize@url \@href}%
\providecommand \@href[1]{\@@startlink{#1}\@@href}%
\providecommand \@@href[1]{\endgroup#1\@@endlink}%
\providecommand \@sanitize@url [0]{\catcode `\\12\catcode `\$12\catcode
  `\&12\catcode `\#12\catcode `\^12\catcode `\_12\catcode `\%12\relax}%
\providecommand \@@startlink[1]{}%
\providecommand \@@endlink[0]{}%
\providecommand \url  [0]{\begingroup\@sanitize@url \@url }%
\providecommand \@url [1]{\endgroup\@href {#1}{\urlprefix }}%
\providecommand \urlprefix  [0]{URL }%
\providecommand \Eprint [0]{\href }%
\providecommand \doibase [0]{http://dx.doi.org/}%
\providecommand \selectlanguage [0]{\@gobble}%
\providecommand \bibinfo  [0]{\@secondoftwo}%
\providecommand \bibfield  [0]{\@secondoftwo}%
\providecommand \translation [1]{[#1]}%
\providecommand \BibitemOpen [0]{}%
\providecommand \bibitemStop [0]{}%
\providecommand \bibitemNoStop [0]{.\EOS\space}%
\providecommand \EOS [0]{\spacefactor3000\relax}%
\providecommand \BibitemShut  [1]{\csname bibitem#1\endcsname}%
\let\auto@bib@innerbib\@empty
\bibitem [{\citenamefont {Klitzing}\ \emph {et~al.}(1980)\citenamefont
  {Klitzing}, \citenamefont {Dorda},\ and\ \citenamefont
  {Pepper}}]{Klitzing1980}%
  \BibitemOpen
  \bibfield  {author} {\bibinfo {author} {\bibfnamefont {K.~v.}\ \bibnamefont
  {Klitzing}}, \bibinfo {author} {\bibfnamefont {G.}~\bibnamefont {Dorda}}, \
  and\ \bibinfo {author} {\bibfnamefont {M.}~\bibnamefont {Pepper}},\
  }\href@noop {} {\bibfield  {journal} {\bibinfo  {journal} {Phys. Rev. Lett.}\
  }\textbf {\bibinfo {volume} {45}},\ \bibinfo {pages} {494} (\bibinfo {year}
  {1980})}\BibitemShut {NoStop}%
\bibitem [{\citenamefont {Onsager}(1949)}]{Onsager1949a}%
  \BibitemOpen
  \bibfield  {author} {\bibinfo {author} {\bibfnamefont {L.}~\bibnamefont
  {Onsager}},\ }\href@noop {} {\bibfield  {journal} {\bibinfo  {journal} {Nuovo
  Cimento}\ }\textbf {\bibinfo {volume} {6}},\ \bibinfo {pages} {249} (\bibinfo
  {year} {1949})}\BibitemShut {NoStop}%
\bibitem [{ang()}]{angular}%
  \BibitemOpen
  \href@noop {} {}\bibinfo {note} {There may also exist an angular shift, i.e.,
  the deviation in the direction of the reflected beam. See e.g., K. Y. Bliokh
  and A. Aiello, J. Opt. \textbf{15}, 014001 (2013).}\BibitemShut {Stop}%
\bibitem [{\citenamefont {Goos}\ and\ \citenamefont
  {H\"{a}nchen}(1947)}]{Goos1947}%
  \BibitemOpen
  \bibfield  {author} {\bibinfo {author} {\bibfnamefont {F.}~\bibnamefont
  {Goos}}\ and\ \bibinfo {author} {\bibfnamefont {H.}~\bibnamefont
  {H\"{a}nchen}},\ }\href@noop {} {\bibfield  {journal} {\bibinfo  {journal}
  {Ann. Phys.}\ }\textbf {\bibinfo {volume} {436}},\ \bibinfo {pages} {333}
  (\bibinfo {year} {1947})}\BibitemShut {NoStop}%
\bibitem [{\citenamefont {Fedorov}(1955)}]{Fedorov1955}%
  \BibitemOpen
  \bibfield  {author} {\bibinfo {author} {\bibfnamefont {F.}~\bibnamefont
  {Fedorov}},\ }\href@noop {} {\bibfield  {journal} {\bibinfo  {journal} {Dokl.
  Akad. Nauk SSSR}\ }\textbf {\bibinfo {volume} {105}},\ \bibinfo {pages} {465}
  (\bibinfo {year} {1955})}\BibitemShut {NoStop}%
\bibitem [{\citenamefont {Imbert}(1972)}]{Imbert1972}%
  \BibitemOpen
  \bibfield  {author} {\bibinfo {author} {\bibfnamefont {C.}~\bibnamefont
  {Imbert}},\ }\href@noop {} {\bibfield  {journal} {\bibinfo  {journal} {Phys.
  Rev. D}\ }\textbf {\bibinfo {volume} {5}},\ \bibinfo {pages} {787} (\bibinfo
  {year} {1972})}\BibitemShut {NoStop}%
\bibitem [{\citenamefont {Miller}\ and\ \citenamefont
  {Ashby}(1972)}]{Miller1972}%
  \BibitemOpen
  \bibfield  {author} {\bibinfo {author} {\bibfnamefont {S.~C.}\ \bibnamefont
  {Miller}}\ and\ \bibinfo {author} {\bibfnamefont {N.}~\bibnamefont {Ashby}},\
  }\href@noop {} {\bibfield  {journal} {\bibinfo  {journal} {Phys. Rev. Lett.}\
  }\textbf {\bibinfo {volume} {29}},\ \bibinfo {pages} {740} (\bibinfo {year}
  {1972})}\BibitemShut {NoStop}%
\bibitem [{\citenamefont {Fradkin}\ and\ \citenamefont
  {Kashuba}(1974)}]{Fradkin1974}%
  \BibitemOpen
  \bibfield  {author} {\bibinfo {author} {\bibfnamefont {D.~M.}\ \bibnamefont
  {Fradkin}}\ and\ \bibinfo {author} {\bibfnamefont {R.~J.}\ \bibnamefont
  {Kashuba}},\ }\href@noop {} {\bibfield  {journal} {\bibinfo  {journal} {Phys.
  Rev. D}\ }\textbf {\bibinfo {volume} {9}},\ \bibinfo {pages} {2775} (\bibinfo
  {year} {1974})}\BibitemShut {NoStop}%
\bibitem [{\citenamefont {Sinitsyn}\ \emph {et~al.}(2005)\citenamefont
  {Sinitsyn}, \citenamefont {Niu}, \citenamefont {Sinova},\ and\ \citenamefont
  {Nomura}}]{Sinitsyn2005}%
  \BibitemOpen
  \bibfield  {author} {\bibinfo {author} {\bibfnamefont {N.~A.}\ \bibnamefont
  {Sinitsyn}}, \bibinfo {author} {\bibfnamefont {Q.}~\bibnamefont {Niu}},
  \bibinfo {author} {\bibfnamefont {J.}~\bibnamefont {Sinova}}, \ and\ \bibinfo
  {author} {\bibfnamefont {K.}~\bibnamefont {Nomura}},\ }\href@noop {}
  {\bibfield  {journal} {\bibinfo  {journal} {Phys. Rev. B}\ }\textbf {\bibinfo
  {volume} {72}},\ \bibinfo {pages} {045346} (\bibinfo {year}
  {2005})}\BibitemShut {NoStop}%
\bibitem [{\citenamefont {Chen}\ \emph {et~al.}(2008)\citenamefont {Chen},
  \citenamefont {Li},\ and\ \citenamefont {Ban}}]{Chen2008}%
  \BibitemOpen
  \bibfield  {author} {\bibinfo {author} {\bibfnamefont {X.}~\bibnamefont
  {Chen}}, \bibinfo {author} {\bibfnamefont {C.-F.}\ \bibnamefont {Li}}, \ and\
  \bibinfo {author} {\bibfnamefont {Y.}~\bibnamefont {Ban}},\ }\href@noop {}
  {\bibfield  {journal} {\bibinfo  {journal} {Phys. Rev. B}\ }\textbf {\bibinfo
  {volume} {77}},\ \bibinfo {pages} {073307} (\bibinfo {year}
  {2008})}\BibitemShut {NoStop}%
\bibitem [{\citenamefont {Beenakker}\ \emph {et~al.}(2009)\citenamefont
  {Beenakker}, \citenamefont {Sepkhanov}, \citenamefont {Akhmerov},\ and\
  \citenamefont {Tworzyd\l{}o}}]{BeenakkerGH2009}%
  \BibitemOpen
  \bibfield  {author} {\bibinfo {author} {\bibfnamefont {C.~W.~J.}\
  \bibnamefont {Beenakker}}, \bibinfo {author} {\bibfnamefont {R.~A.}\
  \bibnamefont {Sepkhanov}}, \bibinfo {author} {\bibfnamefont {A.~R.}\
  \bibnamefont {Akhmerov}}, \ and\ \bibinfo {author} {\bibfnamefont
  {J.}~\bibnamefont {Tworzyd\l{}o}},\ }\href@noop {} {\bibfield  {journal}
  {\bibinfo  {journal} {Phys. Rev. Lett.}\ }\textbf {\bibinfo {volume} {102}},\
  \bibinfo {pages} {146804} (\bibinfo {year} {2009})}\BibitemShut {NoStop}%
\bibitem [{\citenamefont {Sharma}\ and\ \citenamefont
  {Ghosh}(2011)}]{Sharma2011}%
  \BibitemOpen
  \bibfield  {author} {\bibinfo {author} {\bibfnamefont {M.}~\bibnamefont
  {Sharma}}\ and\ \bibinfo {author} {\bibfnamefont {S.}~\bibnamefont {Ghosh}},\
  }\href@noop {} {\bibfield  {journal} {\bibinfo  {journal} {Journal of
  Physics: Condensed Matter}\ }\textbf {\bibinfo {volume} {23}},\ \bibinfo
  {pages} {055501} (\bibinfo {year} {2011})}\BibitemShut {NoStop}%
\bibitem [{\citenamefont {Wu}\ \emph {et~al.}(2011)\citenamefont {Wu},
  \citenamefont {Zhai}, \citenamefont {Peeters}, \citenamefont {Xu},\ and\
  \citenamefont {Chang}}]{WuPRL2011}%
  \BibitemOpen
  \bibfield  {author} {\bibinfo {author} {\bibfnamefont {Z.}~\bibnamefont
  {Wu}}, \bibinfo {author} {\bibfnamefont {F.}~\bibnamefont {Zhai}}, \bibinfo
  {author} {\bibfnamefont {F.~M.}\ \bibnamefont {Peeters}}, \bibinfo {author}
  {\bibfnamefont {H.~Q.}\ \bibnamefont {Xu}}, \ and\ \bibinfo {author}
  {\bibfnamefont {K.}~\bibnamefont {Chang}},\ }\href@noop {} {\bibfield
  {journal} {\bibinfo  {journal} {Phys. Rev. Lett.}\ }\textbf {\bibinfo
  {volume} {106}},\ \bibinfo {pages} {176802} (\bibinfo {year}
  {2011})}\BibitemShut {NoStop}%
\bibitem [{\citenamefont {Chen}\ \emph {et~al.}(2011)\citenamefont {Chen},
  \citenamefont {Tao},\ and\ \citenamefont {Ban}}]{Chen2011}%
  \BibitemOpen
  \bibfield  {author} {\bibinfo {author} {\bibfnamefont {X.}~\bibnamefont
  {Chen}}, \bibinfo {author} {\bibfnamefont {J.-W.}\ \bibnamefont {Tao}}, \
  and\ \bibinfo {author} {\bibfnamefont {Y.}~\bibnamefont {Ban}},\ }\href@noop
  {} {\bibfield  {journal} {\bibinfo  {journal} {The European Physical Journal
  B}\ }\textbf {\bibinfo {volume} {79}},\ \bibinfo {pages} {203} (\bibinfo
  {year} {2011})}\BibitemShut {NoStop}%
\bibitem [{\citenamefont {Jiang}\ \emph {et~al.}(2015)\citenamefont {Jiang},
  \citenamefont {Jiang}, \citenamefont {Liu}, \citenamefont {Sun},\ and\
  \citenamefont {Xie}}]{JiangPRL2015}%
  \BibitemOpen
  \bibfield  {author} {\bibinfo {author} {\bibfnamefont {Q.-D.}\ \bibnamefont
  {Jiang}}, \bibinfo {author} {\bibfnamefont {H.}~\bibnamefont {Jiang}},
  \bibinfo {author} {\bibfnamefont {H.}~\bibnamefont {Liu}}, \bibinfo {author}
  {\bibfnamefont {Q.-F.}\ \bibnamefont {Sun}}, \ and\ \bibinfo {author}
  {\bibfnamefont {X.~C.}\ \bibnamefont {Xie}},\ }\href@noop {} {\bibfield
  {journal} {\bibinfo  {journal} {Phys. Rev. Lett.}\ }\textbf {\bibinfo
  {volume} {115}},\ \bibinfo {pages} {156602} (\bibinfo {year}
  {2015})}\BibitemShut {NoStop}%
\bibitem [{\citenamefont {Yang}\ \emph {et~al.}(2015)\citenamefont {Yang},
  \citenamefont {Pan},\ and\ \citenamefont {Zhang}}]{YangPRL2015}%
  \BibitemOpen
  \bibfield  {author} {\bibinfo {author} {\bibfnamefont {S.~A.}\ \bibnamefont
  {Yang}}, \bibinfo {author} {\bibfnamefont {H.}~\bibnamefont {Pan}}, \ and\
  \bibinfo {author} {\bibfnamefont {F.}~\bibnamefont {Zhang}},\ }\href@noop {}
  {\bibfield  {journal} {\bibinfo  {journal} {Phys. Rev. Lett.}\ }\textbf
  {\bibinfo {volume} {115}},\ \bibinfo {pages} {156603} (\bibinfo {year}
  {2015})}\BibitemShut {NoStop}%
\bibitem [{\citenamefont {Jiang}\ \emph {et~al.}(2016)\citenamefont {Jiang},
  \citenamefont {Jiang}, \citenamefont {Liu}, \citenamefont {Sun},\ and\
  \citenamefont {Xie}}]{JiangPRB2016}%
  \BibitemOpen
  \bibfield  {author} {\bibinfo {author} {\bibfnamefont {Q.-D.}\ \bibnamefont
  {Jiang}}, \bibinfo {author} {\bibfnamefont {H.}~\bibnamefont {Jiang}},
  \bibinfo {author} {\bibfnamefont {H.}~\bibnamefont {Liu}}, \bibinfo {author}
  {\bibfnamefont {Q.-F.}\ \bibnamefont {Sun}}, \ and\ \bibinfo {author}
  {\bibfnamefont {X.~C.}\ \bibnamefont {Xie}},\ }\href@noop {} {\bibfield
  {journal} {\bibinfo  {journal} {Phys. Rev. B}\ }\textbf {\bibinfo {volume}
  {93}},\ \bibinfo {pages} {195165} (\bibinfo {year} {2016})}\BibitemShut
  {NoStop}%
\bibitem [{\citenamefont {Wang}\ and\ \citenamefont
  {Jian}(2017)}]{WangPRB2017}%
  \BibitemOpen
  \bibfield  {author} {\bibinfo {author} {\bibfnamefont {L.}~\bibnamefont
  {Wang}}\ and\ \bibinfo {author} {\bibfnamefont {S.-K.}\ \bibnamefont
  {Jian}},\ }\href@noop {} {\bibfield  {journal} {\bibinfo  {journal} {Phys.
  Rev. B}\ }\textbf {\bibinfo {volume} {96}},\ \bibinfo {pages} {115448}
  (\bibinfo {year} {2017})}\BibitemShut {NoStop}%
\bibitem [{\citenamefont {Chattopadhyay}\ \emph {et~al.}(2019)\citenamefont
  {Chattopadhyay}, \citenamefont {Shi}, \citenamefont {Zhang}, \citenamefont
  {Song},\ and\ \citenamefont {Chong}}]{Chattopadhyay2018}%
  \BibitemOpen
  \bibfield  {author} {\bibinfo {author} {\bibfnamefont {U.}~\bibnamefont
  {Chattopadhyay}}, \bibinfo {author} {\bibfnamefont {L.-k.}\ \bibnamefont
  {Shi}}, \bibinfo {author} {\bibfnamefont {B.}~\bibnamefont {Zhang}}, \bibinfo
  {author} {\bibfnamefont {J.~C.~W.}\ \bibnamefont {Song}}, \ and\ \bibinfo
  {author} {\bibfnamefont {Y.~D.}\ \bibnamefont {Chong}},\ }\href@noop {}
  {\bibfield  {journal} {\bibinfo  {journal} {Phys. Rev. Lett.}\ }\textbf
  {\bibinfo {volume} {122}},\ \bibinfo {pages} {066602} (\bibinfo {year}
  {2019})}\BibitemShut {NoStop}%
\bibitem [{\citenamefont {Liu}\ \emph {et~al.}(2017)\citenamefont {Liu},
  \citenamefont {Yu},\ and\ \citenamefont {Yang}}]{LiuSUTDPRB2017}%
  \BibitemOpen
  \bibfield  {author} {\bibinfo {author} {\bibfnamefont {Y.}~\bibnamefont
  {Liu}}, \bibinfo {author} {\bibfnamefont {Z.-M.}\ \bibnamefont {Yu}}, \ and\
  \bibinfo {author} {\bibfnamefont {S.~A.}\ \bibnamefont {Yang}},\ }\href@noop
  {} {\bibfield  {journal} {\bibinfo  {journal} {Phys. Rev. B}\ }\textbf
  {\bibinfo {volume} {96}},\ \bibinfo {pages} {121101} (\bibinfo {year}
  {2017})}\BibitemShut {NoStop}%
\bibitem [{\citenamefont {Liu}\ \emph {et~al.}(2018{\natexlab{a}})\citenamefont
  {Liu}, \citenamefont {Yu}, \citenamefont {Liu}, \citenamefont {Jiang},\ and\
  \citenamefont {Yang}}]{LiuCAR2018}%
  \BibitemOpen
  \bibfield  {author} {\bibinfo {author} {\bibfnamefont {Y.}~\bibnamefont
  {Liu}}, \bibinfo {author} {\bibfnamefont {Z.-M.}\ \bibnamefont {Yu}},
  \bibinfo {author} {\bibfnamefont {J.}~\bibnamefont {Liu}}, \bibinfo {author}
  {\bibfnamefont {H.}~\bibnamefont {Jiang}}, \ and\ \bibinfo {author}
  {\bibfnamefont {S.~A.}\ \bibnamefont {Yang}},\ }\href@noop {} {\bibfield
  {journal} {\bibinfo  {journal} {Phys. Rev. B}\ }\textbf {\bibinfo {volume}
  {98}},\ \bibinfo {pages} {195141} (\bibinfo {year}
  {2018}{\natexlab{a}})}\BibitemShut {NoStop}%
\bibitem [{\citenamefont {Yu}\ \emph {et~al.}(2018)\citenamefont {Yu},
  \citenamefont {Liu}, \citenamefont {Yao},\ and\ \citenamefont
  {Yang}}]{YuSUTDPRL2018}%
  \BibitemOpen
  \bibfield  {author} {\bibinfo {author} {\bibfnamefont {Z.-M.}\ \bibnamefont
  {Yu}}, \bibinfo {author} {\bibfnamefont {Y.}~\bibnamefont {Liu}}, \bibinfo
  {author} {\bibfnamefont {Y.}~\bibnamefont {Yao}}, \ and\ \bibinfo {author}
  {\bibfnamefont {S.~A.}\ \bibnamefont {Yang}},\ }\href@noop {} {\bibfield
  {journal} {\bibinfo  {journal} {Phys. Rev. Lett.}\ }\textbf {\bibinfo
  {volume} {121}},\ \bibinfo {pages} {176602} (\bibinfo {year}
  {2018})}\BibitemShut {NoStop}%
\bibitem [{\citenamefont {Onoda}\ \emph {et~al.}(2004)\citenamefont {Onoda},
  \citenamefont {Murakami},\ and\ \citenamefont {Nagaosa}}]{OnodaPRL2004}%
  \BibitemOpen
  \bibfield  {author} {\bibinfo {author} {\bibfnamefont {M.}~\bibnamefont
  {Onoda}}, \bibinfo {author} {\bibfnamefont {S.}~\bibnamefont {Murakami}}, \
  and\ \bibinfo {author} {\bibfnamefont {N.}~\bibnamefont {Nagaosa}},\
  }\href@noop {} {\bibfield  {journal} {\bibinfo  {journal} {Phys. Rev. Lett.}\
  }\textbf {\bibinfo {volume} {93}},\ \bibinfo {pages} {083901} (\bibinfo
  {year} {2004})}\BibitemShut {NoStop}%
\bibitem [{\citenamefont {Bliokh}\ and\ \citenamefont
  {Bliokh}(2006)}]{BliokhPRL2006}%
  \BibitemOpen
  \bibfield  {author} {\bibinfo {author} {\bibfnamefont {K.~Y.}\ \bibnamefont
  {Bliokh}}\ and\ \bibinfo {author} {\bibfnamefont {Y.~P.}\ \bibnamefont
  {Bliokh}},\ }\href@noop {} {\bibfield  {journal} {\bibinfo  {journal} {Phys.
  Rev. Lett.}\ }\textbf {\bibinfo {volume} {96}},\ \bibinfo {pages} {073903}
  (\bibinfo {year} {2006})}\BibitemShut {NoStop}%
\bibitem [{\citenamefont {Bliokh}\ \emph {et~al.}(2015)\citenamefont {Bliokh},
  \citenamefont {Rodríguez-Fortuño}, \citenamefont {Nori},\ and\
  \citenamefont {Zayats}}]{BliokhNPhoto2015}%
  \BibitemOpen
  \bibfield  {author} {\bibinfo {author} {\bibfnamefont {K.~Y.}\ \bibnamefont
  {Bliokh}}, \bibinfo {author} {\bibfnamefont {F.~J.}\ \bibnamefont
  {Rodríguez-Fortuño}}, \bibinfo {author} {\bibfnamefont {F.}~\bibnamefont
  {Nori}}, \ and\ \bibinfo {author} {\bibfnamefont {A.~V.}\ \bibnamefont
  {Zayats}},\ }\href@noop {} {\bibfield  {journal} {\bibinfo  {journal} {Nature
  Photonics}\ }\textbf {\bibinfo {volume} {9}},\ \bibinfo {pages} {796}
  (\bibinfo {year} {2015})}\BibitemShut {NoStop}%
\bibitem [{\citenamefont {Kort-Kamp}\ \emph {et~al.}(2016)\citenamefont
  {Kort-Kamp}, \citenamefont {Sinitsyn},\ and\ \citenamefont
  {Dalvit}}]{KKWJMprb2016}%
  \BibitemOpen
  \bibfield  {author} {\bibinfo {author} {\bibfnamefont {W.~J.~M.}\
  \bibnamefont {Kort-Kamp}}, \bibinfo {author} {\bibfnamefont {N.~A.}\
  \bibnamefont {Sinitsyn}}, \ and\ \bibinfo {author} {\bibfnamefont {D.~A.~R.}\
  \bibnamefont {Dalvit}},\ }\href@noop {} {\bibfield  {journal} {\bibinfo
  {journal} {Phys. Rev. B}\ }\textbf {\bibinfo {volume} {93}},\ \bibinfo
  {pages} {081410} (\bibinfo {year} {2016})}\BibitemShut {NoStop}%
\bibitem [{\citenamefont {Yu}\ \emph {et~al.}(2019)\citenamefont {Yu},
  \citenamefont {Liu},\ and\ \citenamefont {Yang}}]{YUFOP2019}%
  \BibitemOpen
  \bibfield  {author} {\bibinfo {author} {\bibfnamefont {Z.-M.}\ \bibnamefont
  {Yu}}, \bibinfo {author} {\bibfnamefont {Y.}~\bibnamefont {Liu}}, \ and\
  \bibinfo {author} {\bibfnamefont {S.~A.}\ \bibnamefont {Yang}},\ }\href@noop
  {} {\bibfield  {journal} {\bibinfo  {journal} {Frontiers of Physics}\
  }\textbf {\bibinfo {volume} {14}},\ \bibinfo {pages} {33402} (\bibinfo {year}
  {2019})}\BibitemShut {NoStop}%
\bibitem [{\citenamefont {Shi}\ and\ \citenamefont {Song}(2019)}]{ShiPRB2019}%
  \BibitemOpen
  \bibfield  {author} {\bibinfo {author} {\bibfnamefont {L.-k.}\ \bibnamefont
  {Shi}}\ and\ \bibinfo {author} {\bibfnamefont {J.~C.~W.}\ \bibnamefont
  {Song}},\ }\href@noop {} {\bibfield  {journal} {\bibinfo  {journal} {Phys.
  Rev. B}\ }\textbf {\bibinfo {volume} {100}},\ \bibinfo {pages} {201405}
  (\bibinfo {year} {2019})}\BibitemShut {NoStop}%
\bibitem [{\citenamefont {Sinitsyn}\ \emph {et~al.}(2006)\citenamefont
  {Sinitsyn}, \citenamefont {Niu},\ and\ \citenamefont
  {MacDonald}}]{SinitsynPRB2006}%
  \BibitemOpen
  \bibfield  {author} {\bibinfo {author} {\bibfnamefont {N.~A.}\ \bibnamefont
  {Sinitsyn}}, \bibinfo {author} {\bibfnamefont {Q.}~\bibnamefont {Niu}}, \
  and\ \bibinfo {author} {\bibfnamefont {A.~H.}\ \bibnamefont {MacDonald}},\
  }\href@noop {} {\bibfield  {journal} {\bibinfo  {journal} {Phys. Rev. B}\
  }\textbf {\bibinfo {volume} {73}},\ \bibinfo {pages} {075318} (\bibinfo
  {year} {2006})}\BibitemShut {NoStop}%
\bibitem [{\citenamefont {Wan}\ \emph {et~al.}(2011)\citenamefont {Wan},
  \citenamefont {Turner}, \citenamefont {Vishwanath},\ and\ \citenamefont
  {Savrasov}}]{Wan2011}%
  \BibitemOpen
  \bibfield  {author} {\bibinfo {author} {\bibfnamefont {X.}~\bibnamefont
  {Wan}}, \bibinfo {author} {\bibfnamefont {A.~M.}\ \bibnamefont {Turner}},
  \bibinfo {author} {\bibfnamefont {A.}~\bibnamefont {Vishwanath}}, \ and\
  \bibinfo {author} {\bibfnamefont {S.~Y.}\ \bibnamefont {Savrasov}},\
  }\href@noop {} {\bibfield  {journal} {\bibinfo  {journal} {Phys. Rev. B}\
  }\textbf {\bibinfo {volume} {83}},\ \bibinfo {pages} {205101} (\bibinfo
  {year} {2011})}\BibitemShut {NoStop}%
\bibitem [{\citenamefont {Armitage}\ \emph {et~al.}(2018)\citenamefont
  {Armitage}, \citenamefont {Mele},\ and\ \citenamefont
  {Vishwanath}}]{ArmitageRMP2018}%
  \BibitemOpen
  \bibfield  {author} {\bibinfo {author} {\bibfnamefont {N.~P.}\ \bibnamefont
  {Armitage}}, \bibinfo {author} {\bibfnamefont {E.~J.}\ \bibnamefont {Mele}},
  \ and\ \bibinfo {author} {\bibfnamefont {A.}~\bibnamefont {Vishwanath}},\
  }\href@noop {} {\bibfield  {journal} {\bibinfo  {journal} {Rev. Mod. Phys.}\
  }\textbf {\bibinfo {volume} {90}},\ \bibinfo {pages} {015001} (\bibinfo
  {year} {2018})}\BibitemShut {NoStop}%
\bibitem [{\citenamefont {Fang}\ \emph {et~al.}(2012)\citenamefont {Fang},
  \citenamefont {Gilbert}, \citenamefont {Dai},\ and\ \citenamefont
  {Bernevig}}]{FangPRL2012}%
  \BibitemOpen
  \bibfield  {author} {\bibinfo {author} {\bibfnamefont {C.}~\bibnamefont
  {Fang}}, \bibinfo {author} {\bibfnamefont {M.~J.}\ \bibnamefont {Gilbert}},
  \bibinfo {author} {\bibfnamefont {X.}~\bibnamefont {Dai}}, \ and\ \bibinfo
  {author} {\bibfnamefont {B.~A.}\ \bibnamefont {Bernevig}},\ }\href@noop {}
  {\bibfield  {journal} {\bibinfo  {journal} {Phys. Rev. Lett.}\ }\textbf
  {\bibinfo {volume} {108}},\ \bibinfo {pages} {266802} (\bibinfo {year}
  {2012})}\BibitemShut {NoStop}%
\bibitem [{\citenamefont {Xu}\ \emph {et~al.}(2011)\citenamefont {Xu},
  \citenamefont {Weng}, \citenamefont {Wang}, \citenamefont {Dai},\ and\
  \citenamefont {Fang}}]{Xu2011}%
  \BibitemOpen
  \bibfield  {author} {\bibinfo {author} {\bibfnamefont {G.}~\bibnamefont
  {Xu}}, \bibinfo {author} {\bibfnamefont {H.}~\bibnamefont {Weng}}, \bibinfo
  {author} {\bibfnamefont {Z.}~\bibnamefont {Wang}}, \bibinfo {author}
  {\bibfnamefont {X.}~\bibnamefont {Dai}}, \ and\ \bibinfo {author}
  {\bibfnamefont {Z.}~\bibnamefont {Fang}},\ }\href@noop {} {\bibfield
  {journal} {\bibinfo  {journal} {Phys. Rev. Lett.}\ }\textbf {\bibinfo
  {volume} {107}},\ \bibinfo {pages} {186806} (\bibinfo {year}
  {2011})}\BibitemShut {NoStop}%
\bibitem [{SI()}]{SI}%
  \BibitemOpen
  \href@noop {} {}\bibinfo {note} {See Supplemental Material for the detailed
  derivation of the shift and the CAS, as well as the influence of interface
  potential barrier on quantized CAS, which includes
  \cite{YuSUTDPRL2018,Blonder1982,BenDaniel1966,JongPRL1995}}\BibitemShut
  {NoStop}%
\bibitem [{\citenamefont {de~Gennes}(1966)}]{Gennes1966}%
  \BibitemOpen
  \bibfield  {author} {\bibinfo {author} {\bibfnamefont {P.~G.}\ \bibnamefont
  {de~Gennes}},\ }\href@noop {} {\emph {\bibinfo {title} {Superconductivity in
  Metals and Alloys}}}\ (\bibinfo  {publisher} {Benjamin},\ \bibinfo {address}
  {New York},\ \bibinfo {year} {1966})\BibitemShut {NoStop}%
\bibitem [{\citenamefont {Blonder}\ \emph {et~al.}(1982)\citenamefont
  {Blonder}, \citenamefont {Tinkham},\ and\ \citenamefont
  {Klapwijk}}]{Blonder1982}%
  \BibitemOpen
  \bibfield  {author} {\bibinfo {author} {\bibfnamefont {G.~E.}\ \bibnamefont
  {Blonder}}, \bibinfo {author} {\bibfnamefont {M.}~\bibnamefont {Tinkham}}, \
  and\ \bibinfo {author} {\bibfnamefont {T.~M.}\ \bibnamefont {Klapwijk}},\
  }\href@noop {} {\bibfield  {journal} {\bibinfo  {journal} {Phys. Rev. B}\
  }\textbf {\bibinfo {volume} {25}},\ \bibinfo {pages} {4515} (\bibinfo {year}
  {1982})}\BibitemShut {NoStop}%
\bibitem [{\citenamefont {Kashiwaya}\ and\ \citenamefont
  {Tanaka}(2000)}]{Kashiwaya2000}%
  \BibitemOpen
  \bibfield  {author} {\bibinfo {author} {\bibfnamefont {S.}~\bibnamefont
  {Kashiwaya}}\ and\ \bibinfo {author} {\bibfnamefont {Y.}~\bibnamefont
  {Tanaka}},\ }\href@noop {} {\bibfield  {journal} {\bibinfo  {journal}
  {Reports on Progress in Physics}\ }\textbf {\bibinfo {volume} {63}},\
  \bibinfo {pages} {1641} (\bibinfo {year} {2000})}\BibitemShut {NoStop}%
\bibitem [{\citenamefont {Hosten}\ and\ \citenamefont
  {Kwiat}(2008)}]{Hosten2008}%
  \BibitemOpen
  \bibfield  {author} {\bibinfo {author} {\bibfnamefont {O.}~\bibnamefont
  {Hosten}}\ and\ \bibinfo {author} {\bibfnamefont {P.}~\bibnamefont {Kwiat}},\
  }\href@noop {} {\bibfield  {journal} {\bibinfo  {journal} {Science}\ }\textbf
  {\bibinfo {volume} {319}},\ \bibinfo {pages} {787} (\bibinfo {year}
  {2008})}\BibitemShut {NoStop}%
\bibitem [{\citenamefont {Bliokh}\ \emph {et~al.}(2008)\citenamefont {Bliokh},
  \citenamefont {Niv}, \citenamefont {Kleiner},\ and\ \citenamefont
  {Hasman}}]{BliokhNPhoto2008}%
  \BibitemOpen
  \bibfield  {author} {\bibinfo {author} {\bibfnamefont {K.~Y.}\ \bibnamefont
  {Bliokh}}, \bibinfo {author} {\bibfnamefont {A.}~\bibnamefont {Niv}},
  \bibinfo {author} {\bibfnamefont {V.}~\bibnamefont {Kleiner}}, \ and\
  \bibinfo {author} {\bibfnamefont {E.}~\bibnamefont {Hasman}},\ }\href@noop {}
  {\bibfield  {journal} {\bibinfo  {journal} {Nature Photonics}\ }\textbf
  {\bibinfo {volume} {2}},\ \bibinfo {pages} {748} (\bibinfo {year}
  {2008})}\BibitemShut {NoStop}%
\bibitem [{\citenamefont {Haefner}\ \emph {et~al.}(2009)\citenamefont
  {Haefner}, \citenamefont {Sukhov},\ and\ \citenamefont
  {Dogariu}}]{Haefner2009}%
  \BibitemOpen
  \bibfield  {author} {\bibinfo {author} {\bibfnamefont {D.}~\bibnamefont
  {Haefner}}, \bibinfo {author} {\bibfnamefont {S.}~\bibnamefont {Sukhov}}, \
  and\ \bibinfo {author} {\bibfnamefont {A.}~\bibnamefont {Dogariu}},\
  }\href@noop {} {\bibfield  {journal} {\bibinfo  {journal} {Phys. Rev. Lett.}\
  }\textbf {\bibinfo {volume} {102}},\ \bibinfo {pages} {123903} (\bibinfo
  {year} {2009})}\BibitemShut {NoStop}%
\bibitem [{\citenamefont {Rodr\'{\i}guez-Herrera}\ \emph
  {et~al.}(2010)\citenamefont {Rodr\'{\i}guez-Herrera}, \citenamefont {Lara},
  \citenamefont {Bliokh}, \citenamefont {Ostrovskaya},\ and\ \citenamefont
  {Dainty}}]{Herrera2010}%
  \BibitemOpen
  \bibfield  {author} {\bibinfo {author} {\bibfnamefont {O.~G.}\ \bibnamefont
  {Rodr\'{\i}guez-Herrera}}, \bibinfo {author} {\bibfnamefont {D.}~\bibnamefont
  {Lara}}, \bibinfo {author} {\bibfnamefont {K.~Y.}\ \bibnamefont {Bliokh}},
  \bibinfo {author} {\bibfnamefont {E.~A.}\ \bibnamefont {Ostrovskaya}}, \ and\
  \bibinfo {author} {\bibfnamefont {C.}~\bibnamefont {Dainty}},\ }\href@noop {}
  {\bibfield  {journal} {\bibinfo  {journal} {Phys. Rev. Lett.}\ }\textbf
  {\bibinfo {volume} {104}},\ \bibinfo {pages} {253601} (\bibinfo {year}
  {2010})}\BibitemShut {NoStop}%
\bibitem [{\citenamefont {Yin}\ \emph {et~al.}(2013)\citenamefont {Yin},
  \citenamefont {Ye}, \citenamefont {Rho}, \citenamefont {Wang},\ and\
  \citenamefont {Zhang}}]{Yin2013}%
  \BibitemOpen
  \bibfield  {author} {\bibinfo {author} {\bibfnamefont {X.}~\bibnamefont
  {Yin}}, \bibinfo {author} {\bibfnamefont {Z.}~\bibnamefont {Ye}}, \bibinfo
  {author} {\bibfnamefont {J.}~\bibnamefont {Rho}}, \bibinfo {author}
  {\bibfnamefont {Y.}~\bibnamefont {Wang}}, \ and\ \bibinfo {author}
  {\bibfnamefont {X.}~\bibnamefont {Zhang}},\ }\href@noop {} {\bibfield
  {journal} {\bibinfo  {journal} {Science}\ }\textbf {\bibinfo {volume}
  {339}},\ \bibinfo {pages} {1405} (\bibinfo {year} {2013})}\BibitemShut
  {NoStop}%
\bibitem [{\citenamefont {Zhou}\ \emph {et~al.}(2013)\citenamefont {Zhou},
  \citenamefont {Zhang}, \citenamefont {Ling}, \citenamefont {Chen},
  \citenamefont {Luo},\ and\ \citenamefont {Wen}}]{ZhouPRA2013}%
  \BibitemOpen
  \bibfield  {author} {\bibinfo {author} {\bibfnamefont {X.}~\bibnamefont
  {Zhou}}, \bibinfo {author} {\bibfnamefont {J.}~\bibnamefont {Zhang}},
  \bibinfo {author} {\bibfnamefont {X.}~\bibnamefont {Ling}}, \bibinfo {author}
  {\bibfnamefont {S.}~\bibnamefont {Chen}}, \bibinfo {author} {\bibfnamefont
  {H.}~\bibnamefont {Luo}}, \ and\ \bibinfo {author} {\bibfnamefont
  {S.}~\bibnamefont {Wen}},\ }\href@noop {} {\bibfield  {journal} {\bibinfo
  {journal} {Phys. Rev. A}\ }\textbf {\bibinfo {volume} {88}},\ \bibinfo
  {pages} {053840} (\bibinfo {year} {2013})}\BibitemShut {NoStop}%
\bibitem [{\citenamefont {Lu}\ \emph {et~al.}(2015)\citenamefont {Lu},
  \citenamefont {Wang}, \citenamefont {Ye}, \citenamefont {Ran}, \citenamefont
  {Fu}, \citenamefont {Joannopoulos},\ and\ \citenamefont {Solja{\v
  c}i{\'c}}}]{Lu6222015}%
  \BibitemOpen
  \bibfield  {author} {\bibinfo {author} {\bibfnamefont {L.}~\bibnamefont
  {Lu}}, \bibinfo {author} {\bibfnamefont {Z.}~\bibnamefont {Wang}}, \bibinfo
  {author} {\bibfnamefont {D.}~\bibnamefont {Ye}}, \bibinfo {author}
  {\bibfnamefont {L.}~\bibnamefont {Ran}}, \bibinfo {author} {\bibfnamefont
  {L.}~\bibnamefont {Fu}}, \bibinfo {author} {\bibfnamefont {J.~D.}\
  \bibnamefont {Joannopoulos}}, \ and\ \bibinfo {author} {\bibfnamefont
  {M.}~\bibnamefont {Solja{\v c}i{\'c}}},\ }\href@noop {} {\bibfield  {journal}
  {\bibinfo  {journal} {Science}\ }\textbf {\bibinfo {volume} {349}},\ \bibinfo
  {pages} {622} (\bibinfo {year} {2015})}\BibitemShut {NoStop}%
\bibitem [{\citenamefont {Yang}\ \emph {et~al.}(2018)\citenamefont {Yang},
  \citenamefont {Guo}, \citenamefont {Tremain}, \citenamefont {Liu},
  \citenamefont {Barr}, \citenamefont {Yan}, \citenamefont {Gao}, \citenamefont
  {Liu}, \citenamefont {Xiang}, \citenamefont {Chen}, \citenamefont {Fang},
  \citenamefont {Hibbins}, \citenamefont {Lu},\ and\ \citenamefont
  {Zhang}}]{Yang1013Sci}%
  \BibitemOpen
  \bibfield  {author} {\bibinfo {author} {\bibfnamefont {B.}~\bibnamefont
  {Yang}}, \bibinfo {author} {\bibfnamefont {Q.}~\bibnamefont {Guo}}, \bibinfo
  {author} {\bibfnamefont {B.}~\bibnamefont {Tremain}}, \bibinfo {author}
  {\bibfnamefont {R.}~\bibnamefont {Liu}}, \bibinfo {author} {\bibfnamefont
  {L.~E.}\ \bibnamefont {Barr}}, \bibinfo {author} {\bibfnamefont
  {Q.}~\bibnamefont {Yan}}, \bibinfo {author} {\bibfnamefont {W.}~\bibnamefont
  {Gao}}, \bibinfo {author} {\bibfnamefont {H.}~\bibnamefont {Liu}}, \bibinfo
  {author} {\bibfnamefont {Y.}~\bibnamefont {Xiang}}, \bibinfo {author}
  {\bibfnamefont {J.}~\bibnamefont {Chen}}, \bibinfo {author} {\bibfnamefont
  {C.}~\bibnamefont {Fang}}, \bibinfo {author} {\bibfnamefont {A.}~\bibnamefont
  {Hibbins}}, \bibinfo {author} {\bibfnamefont {L.}~\bibnamefont {Lu}}, \ and\
  \bibinfo {author} {\bibfnamefont {S.}~\bibnamefont {Zhang}},\ }\href@noop {}
  {\bibfield  {journal} {\bibinfo  {journal} {Science}\ }\textbf {\bibinfo
  {volume} {359}},\ \bibinfo {pages} {1013} (\bibinfo {year}
  {2018})}\BibitemShut {NoStop}%
\bibitem [{\citenamefont {Xue}\ \emph {et~al.}(2019)\citenamefont {Xue},
  \citenamefont {Yang}, \citenamefont {Gao}, \citenamefont {Chong},\ and\
  \citenamefont {Zhang}}]{XueNM2019}%
  \BibitemOpen
  \bibfield  {author} {\bibinfo {author} {\bibfnamefont {H.}~\bibnamefont
  {Xue}}, \bibinfo {author} {\bibfnamefont {Y.}~\bibnamefont {Yang}}, \bibinfo
  {author} {\bibfnamefont {F.}~\bibnamefont {Gao}}, \bibinfo {author}
  {\bibfnamefont {Y.}~\bibnamefont {Chong}}, \ and\ \bibinfo {author}
  {\bibfnamefont {B.}~\bibnamefont {Zhang}},\ }\href@noop {} {\bibfield
  {journal} {\bibinfo  {journal} {Nature Materials}\ }\textbf {\bibinfo
  {volume} {18}},\ \bibinfo {pages} {108} (\bibinfo {year} {2019})}\BibitemShut
  {NoStop}%
\bibitem [{\citenamefont {Yang}\ \emph {et~al.}(2019)\citenamefont {Yang},
  \citenamefont {Gao}, \citenamefont {Xue}, \citenamefont {Zhang},
  \citenamefont {He}, \citenamefont {Yang}, \citenamefont {Singh},
  \citenamefont {Chong}, \citenamefont {Zhang},\ and\ \citenamefont
  {Chen}}]{YangNP2019}%
  \BibitemOpen
  \bibfield  {author} {\bibinfo {author} {\bibfnamefont {Y.}~\bibnamefont
  {Yang}}, \bibinfo {author} {\bibfnamefont {Z.}~\bibnamefont {Gao}}, \bibinfo
  {author} {\bibfnamefont {H.}~\bibnamefont {Xue}}, \bibinfo {author}
  {\bibfnamefont {L.}~\bibnamefont {Zhang}}, \bibinfo {author} {\bibfnamefont
  {M.}~\bibnamefont {He}}, \bibinfo {author} {\bibfnamefont {Z.}~\bibnamefont
  {Yang}}, \bibinfo {author} {\bibfnamefont {R.}~\bibnamefont {Singh}},
  \bibinfo {author} {\bibfnamefont {Y.}~\bibnamefont {Chong}}, \bibinfo
  {author} {\bibfnamefont {B.}~\bibnamefont {Zhang}}, \ and\ \bibinfo {author}
  {\bibfnamefont {H.}~\bibnamefont {Chen}},\ }\href@noop {} {\bibfield
  {journal} {\bibinfo  {journal} {Nature}\ }\textbf {\bibinfo {volume} {565}},\
  \bibinfo {pages} {622} (\bibinfo {year} {2019})}\BibitemShut {NoStop}%
\bibitem [{\citenamefont {Liu}\ \emph {et~al.}(2018{\natexlab{b}})\citenamefont
  {Liu}, \citenamefont {Yu}, \citenamefont {Jiang},\ and\ \citenamefont
  {Yang}}]{LiuSUTD2018}%
  \BibitemOpen
  \bibfield  {author} {\bibinfo {author} {\bibfnamefont {Y.}~\bibnamefont
  {Liu}}, \bibinfo {author} {\bibfnamefont {Z.-M.}\ \bibnamefont {Yu}},
  \bibinfo {author} {\bibfnamefont {H.}~\bibnamefont {Jiang}}, \ and\ \bibinfo
  {author} {\bibfnamefont {S.~A.}\ \bibnamefont {Yang}},\ }\href@noop {}
  {\bibfield  {journal} {\bibinfo  {journal} {Phys. Rev. B}\ }\textbf {\bibinfo
  {volume} {98}},\ \bibinfo {pages} {075151} (\bibinfo {year}
  {2018}{\natexlab{b}})}\BibitemShut {NoStop}%
\bibitem [{\citenamefont {BenDaniel}\ and\ \citenamefont
  {Duke}(1966)}]{BenDaniel1966}%
  \BibitemOpen
  \bibfield  {author} {\bibinfo {author} {\bibfnamefont {D.~J.}\ \bibnamefont
  {BenDaniel}}\ and\ \bibinfo {author} {\bibfnamefont {C.~B.}\ \bibnamefont
  {Duke}},\ }\href@noop {} {\bibfield  {journal} {\bibinfo  {journal} {Phys.
  Rev.}\ }\textbf {\bibinfo {volume} {152}},\ \bibinfo {pages} {683} (\bibinfo
  {year} {1966})}\BibitemShut {NoStop}%
\bibitem [{\citenamefont {de~Jong}\ and\ \citenamefont
  {Beenakker}(1995)}]{JongPRL1995}%
  \BibitemOpen
  \bibfield  {author} {\bibinfo {author} {\bibfnamefont {M.~J.~M.}\
  \bibnamefont {de~Jong}}\ and\ \bibinfo {author} {\bibfnamefont {C.~W.~J.}\
  \bibnamefont {Beenakker}},\ }\href@noop {} {\bibfield  {journal} {\bibinfo
  {journal} {Phys. Rev. Lett.}\ }\textbf {\bibinfo {volume} {74}},\ \bibinfo
  {pages} {1657} (\bibinfo {year} {1995})}\BibitemShut {NoStop}%
\end{thebibliography}%

\end{document}